\documentclass[preprint]{aastex63}
\usepackage{amsmath}
\shorttitle{Pulsar Polarization Estimation}
\shortauthors{McKINNON}

\begin{document}
\title{Polarization Estimation for Radio Pulsars}
\author{M. M. McKinnon}
\affiliation{National Radio Astronomy Observatory, Socorro, NM \ 87801 \ \ USA}
\correspondingauthor{M. M. McKinnon}
\email{mmckinno@nrao.edu}


\begin{abstract}

A number of polarization estimators have been developed for a variety of astrophysical
applications to compensate measurements of linear polarization for a bias contributed by 
the instrumental noise. Most derivations of the estimators assume that the amplitude 
and orientation of the polarization vector are constant. This assumption generally is 
not valid for the radio emission from pulsars that fluctuates from pulse to pulse. The 
radio emission from pulsars, fast radio bursts, and magnetars can be elliptically 
polarized, and estimators of the total polarization and absolute value of the circular 
polarization are used in their observations. However, these estimators have not been 
formally developed to a level that is commensurate with those of linear polarization. 
Estimators are derived for circular, linear, and total polarization when the amplitude 
of the polarization vector is a constant or a random variable. Hybrid estimators are 
proposed for general application to pulsar polarization observations. They are shown 
to be more effective at removing instrumental noise than their commonly used counterparts.

\end{abstract}


\section{Introduction}
\label{sec:intro}

\subsection{Classical Estimators of Linear Polarization}

Measurements of linear polarization contain a contribution from the instrumental noise, 
and a variety of methods have been developed to remove this instrumental bias from the
measurements. The methods generally assume that the amplitude and orientation of the 
polarization vector are constant and the instrumental noise in each of the Stokes parameters 
$Q$ and $U$ is a Gaussian random variable (RV) such that the measured linear polarization 
($L=\sqrt{Q^2+U^2}$) follows a Rice probability density function (pdf). Most methods use 
this pdf to derive a polarization estimator from which the intrinsic polarization may be 
inferred. Serkowski (1958) suggested that the measured polarization is equal to the mean of 
the pdf, while Wardle \& Kronberg (1974) proposed that the measured polarization is equal 
to the mode of the pdf. Simmons \& Stewart (1985) noted that the measured polarization 
could also be represented by the median and maximum likelihood (ML) of the pdf. They showed 
that all four estimators converge to the same result when the signal-to-noise ratio (SNR) in 
polarization is large. The relationship between the measured linear polarization, $L_m$, and 
the intrinsic polarization, $\mu_q$, in this SNR regime is $L_m=\sqrt{\mu_q^2+\sigma_n^2}$, 
where $\sigma_n$ is the magnitude of the instrumental noise. The convergence of the 
estimators arises from the Rice pdf evolving to a Gaussian pdf with a mean of 
$\sqrt{\mu_q^2+\sigma_n^2}$ and a standard deviation of $\sigma_n$, and because the mean, 
mode, median, and ML of a Gaussian pdf are equal to one another. The asymptotic behavior 
of the estimators is often cited as the asymptotic, or AS, estimator (e.g., Plaszczynski 
et al., 2014; Montier et al., 2015). 

\subsection{Estimator Approximations}

The mathematical expressions for the estimators can be complex and are not easily 
implemented in the data analysis. Instead, approximations are often made to the 
estimators using (e.g., Equation 16 of Simmons \& Stewart 1985)
\begin{equation}
L_t =
\begin{cases}
  \left(L_m^2 - K_a^2\sigma_n^2\right)^{1/2} & \quad \mathrm{if}\ L_m > K_a\sigma_n \\
  0 & \quad \mathrm{otherwise},
\end{cases}
\label{eqn:approx}
\end{equation}
where $L_t$ is an estimate of the true polarization and $K_a$ is a threshold value
determined by each estimator. The AS estimator is Equation~\ref{eqn:approx} with $K_a=1$ 
and is equivalent to the approximation to Wardle \& Kronberg's (1974) mode estimator. The 
pdf of the values of $L_t$ resulting from the application of Equation~\ref{eqn:approx} 
is discontinuous. It is comprised of a delta function at $L_t=0$, owing to the cutoff at 
$K_a\sigma_n$, and a truncated Gaussian-like function (see, e.g., Figure 1 of Montier et al., 
2015). The amplitude of the delta function increases with increasing $K_a$. If the value 
of $K_a$ is too large, the measurements are overcompensated for instrumental noise, and 
the estimated value of the true polarization is systematically low. Conversely, if $K_a$ 
is too small, the measurements are undercompensated for noise, and the polarization 
estimate is systematically high. 

\subsection{Estimator Residual Bias and Risk}

Although the classical estimators are effective in removing most of the instrumental 
bias, a residual bias remains after their application. Simmons \& Stewart (1985) 
compared the effectiveness of the mode, mean, median, and ML estimators by calculating 
their residual bias and risk, or standard error about the true polarization, from the pdf
of $L_t$. The residual bias, $B$, and risk, $R$, are defined by
\begin{equation}
B = (\langle L_t\rangle-\mu_q)/\sigma_n,
\end{equation}
\begin{equation}
R = \langle(L_t-\mu_q)^2\rangle/\sigma_n^2,
\end{equation}
where the angular brackets denote averages over the pdf. They found that the ML estimator 
produced the lowest residual bias at low SNR, and the mode estimator produced the lowest 
residual bias at intermediate to high SNR. Although not attainable in practice 
(Plaszczynski et al., 2014), a perfect estimator would produce a residual bias of zero 
and a risk of 1 for all values of $\mu_q$.

\subsection{Additional Estimators}

More sophisticated estimators have been developed to measure linear polarization, 
primarily for the cosmic microwave background radiation. Quinn (2012) proposed a 
Bayesian analysis for estimates of the polarization and its confidence intervals. 
Plaszczynski et al. (2014) developed a modified asymptotic (MAS) estimator that 
attenuates low SNR measurements, instead of setting all values below a certain 
threshold equal to zero, as in Equation~\ref{eqn:approx}. The pdf of $L_t$ resulting 
from the MAS estimator is continuous. This estimator is essentially unbiased for 
SNRs greater than about 3. The classical estimators summarized in Simmons \& Stewart 
(1985) assume the instrumental noise in each of the Stokes parameters $Q$ and $U$ is 
statistically independent and equal in magnitude. Plaszczynski et al. (2014) and 
Montier et al. (2015) evaluate the case when the noise in $Q$ and $U$ is covariant 
and not equal.

Another estimator assumes that the measured polarization is the true polarization. 
Stewart \& Simmons (1985) characterized this assumption as naive, a designation that 
has persisted in the literature to note that no attempt has been made to compensate 
the measurements for instrumental bias. The naive estimator is useful for demonstrating 
the improvements that can be made in bias removal with other estimators.

\subsection{Pulsar Applications}

For a number of reasons, additional development of the classical polarization estimators 
is needed for their application to radio pulsars, fast radio bursts (FRBs), and magnetars. 
First, in addition to reproducing the intrinsic polarization at high SNR, the estimators 
must minimize the contribution of the noise at low SNR so that the polarization off the 
pulse or burst is consistent with zero (e.g., Karastergiou \& Johnson 2004). Everett \& 
Weisberg (2001) developed a hybrid estimator, hereafter the EW estimator, for linear 
polarization that employs the AS estimator for bias compensation at high SNR and a cutoff 
of $L_m=1.57\sigma_n$ for bias compensation at low SNR. The EW estimator is commonly used 
in polarization observations of pulsars and FRBs (e.g., Johnston \& Kerr 2018; Day et al., 
2020; Serylak et al., 2021; Johnston et al., 2023; Posselt et al., 2023; Mckinven et al., 
2023; Basu et al., 2024). Second, pulsar radio emission is generally elliptically 
polarized and frequently exhibits modes of orthogonal polarization (or OPMs; e.g., Manchester 
et al., 1975; Cordes et al., 1978; Backer \& Rankin 1980; Stinebring et al., 1984). 
The superposition of incoherent OPMs can alter the polarization fraction across a pulsar's 
average profile as the relative intensity of the modes changes with pulse longitude 
(McKinnon \& Stinebring 1998, 2000). The polarization vectors of pulsars, FRBs, and 
magnetars can also trace arcs or great circles on the Poincar\'e sphere, respectively, as 
functions of pulse longitude (Dyks et al., 2021; Oswald et al., 2023b), time (Bera et al., 
2025), and frequency (Lower et al., 2024). Their circular ($V$), linear ($L$), and total 
($P$) polarization can be low due to vector crossings of the sphere's equator or poles, 
or to the simultaneous occurence of incoherent OPMs of comparable intensity, making their 
measurement susceptible to the effects of instrumental noise. Therefore, accurate 
polarization measurements and applicable polarization estimators for $V$ and $P$, in 
addition to $L$, are needed to test the viability of models developed for pulsar 
polarization (e.g., Dyks 2020; Dyks et al., 2021; Oswald et al., 2023b; McKinnon 2024). 
Karastergiou et al. (2003) and Karastergiou \& Johnston (2004) developed a polarization 
estimator for the absolute value of circular polarization, $|V|$. Quinn (2014) derived 
the mode and ML estimators for $P=\sqrt{Q^2+U^2+V^2}$, and showed that measured values 
of $P$ vary as $P_m=\sqrt{\mu_p^2 + 2\sigma_n^2}$ at high SNR. Apart from these two 
examples, estimators for $|V|$ and $P$ have yet to be developed to a level that is 
commensurate with those of $L$. Finally, the assumption that the amplitude of a 
polarization vector is constant generally is not valid for radio pulsars, because the 
emission's total intensity and polarization at a particular pulse longitude fluctuate 
from pulse to pulse. The modulation index of the total intensity, the ratio of its 
standard deviation to its mean, is indicative of the emission's heavy modulation. For 
example, the 1352 MHz observations of Burke-Spolaor et al. (2012, their Figure 6) show that 
the minimum modulation index, $\beta$, observed across pulse profiles lies in the range 
of about $0.1<\beta<2$ with a median value of $\beta\simeq 1$ (see also Weltevedre et 
al., 2006, 2007). The modulation of the polarization is likely comparable to that of the 
total intensity. The pdf of the total intensity can range from Gaussian at $\beta<0.2$ 
(e.g., McKinnon 2004), to exponential at $\beta=1$, and log-normal for larger values of 
$\beta$ (e.g., Cairns et al., 2001, 2004). 

The fluctuations in polarized intensity raise a number of related questions regarding 
polarization estimation. Are the estimators derived from the assumption of a polarization 
vector with constant amplitude applicable to a vector with randomly varying amplitude? 
Must polarization estimators be customized to the statistical character of the polarization 
fluctuations? Can a single estimator for a specific polarization ($|V|$, $L$, or $P$) be 
developed for general use, regardless of the statistical character of the fluctuations? 
The purpose of this paper is to address these questions by deriving and comparing
polarization estimators for $|V|$, $L$, and $P$ when the amplitude of the polarization 
vector is constant or random. An overarching goal of the paper is to improve estimators 
that are commonly used for pulsar polarization observations.

The paper is organized as follows. In Section~\ref{sec:constant}, mode, median, mean, 
and ML estimators are derived for $|V|$ and $P$ for the case when the amplitude of the 
polarization vector is constant. The derivations assume the instrumental noise in each of 
the Stokes parameters are independent, Gaussian RVs with equal magnitude. A general hybrid 
estimator for $|V|$, $L$, and $P$ that uses the EW estimator as a template is proposed. 
In Section~\ref{sec:random}, polarization estimators are derived for $|V|$, $L$, and $P$ 
when the amplitude of the polarization vector is a Gaussian or exponential RV. The 
parameterization of the hybrid estimators is optimized by minimizing their residual bias. 
In Section~\ref{sec:compare}, the effectiveness of the hybrid estimators in removing the 
instrumental noise is compared with that of their commonly used counterparts. Summary 
comments are listed in Section~\ref{sec:summary}. Appendix~\ref{sec:Pest} lists the 
equations for estimators of $P$ when the amplitude of a polarization vector is constant. 
Appendix~\ref{sec:OPMest} lists the equations for the estimators of $|V|$, $L$, and $P$ 
when the polarization fluctuations are Gaussian.


\section{Estimators for a Polarization Vector with Constant Amplitude}
\label{sec:constant}

\subsection{Circular Polarization Estimation}
\label{sec:Vconst}

\begin{figure}
\plotone{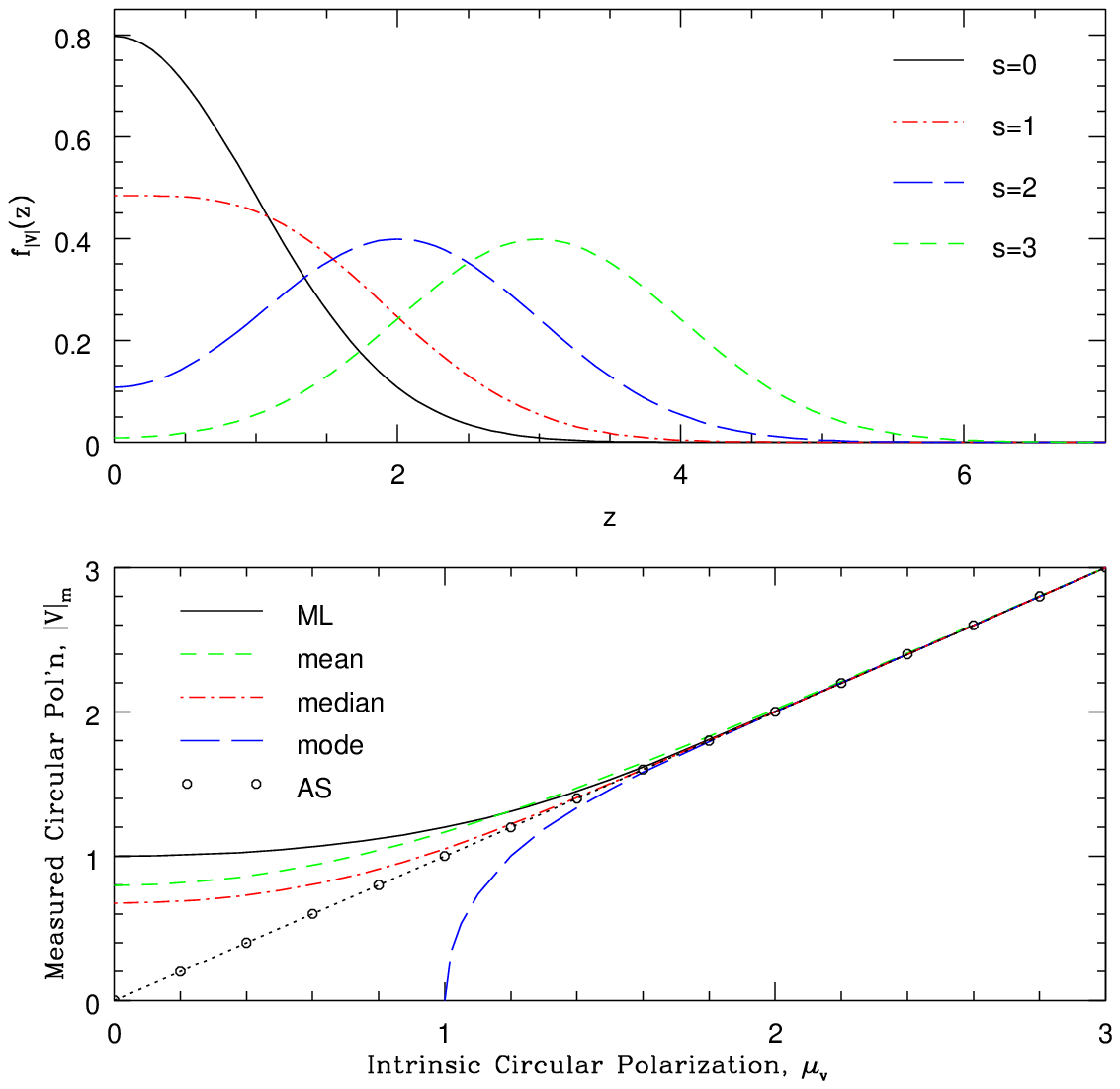}
\caption{Probability density function (pdf) and polarization estimators for the absolute 
value of circular polarization when the amplitude of the polarization vector is constant. 
The top panel shows the pdf of $|V|$ (Equation~\ref{eqn:fv}) for different values of SNR 
in intrinsic circular polarization, $s=\mu_v/\sigma_n$. The bottom panel compares the ML, 
mean, median, mode, and AS estimators for $|V|$. The dotted diagonal line denotes 
$|V|_m=\mu_v$. As with all figures, the magnitude of the instrumental noise used in the 
figure is $\sigma_n=1$.}
\label{fig:Vdist}
\end{figure}

Polarization estimators and their threshold values are derived from the pdf of the measured 
amplitude of the polarization vector. When the intrinsic circular polarization, $V$, is 
constant at $\mu_v$, the pdf of the measured values of $V$ is Gaussian with a mean of $\mu_v$ 
and a standard deviation of $\sigma_n$. The pdf of the absolute value of $V$ is the sum of 
two Gaussian functions (e.g., Equation 27 of McKinnon \& Stinebring 1998):
\begin{equation}
f_{|V|}(z) = \frac{1}{\sigma_n\sqrt{2\pi}}
             \left\{\exp{\left[-\frac{(z+\mu_v)^2}{2\sigma_n^2}\right]} +
             \exp{\left[-\frac{(z-\mu_v)^2}{2\sigma_n^2}\right]}\right\}, \qquad z\ge 0.
\label{eqn:fv}
\end{equation}
Examples of $f_{|V|}(z)$ are shown in the top panel of Figure~\ref{fig:Vdist} for different 
SNR values in intrinsic circular polarization, $s=\mu_v/\sigma_n$. 

The mean estimator for $|V|$ is the mean of $f_{|V|}(z)$ (Serkowski 1958) and is given by 
(e.g., Equation 13 of McKinnon \& Stinebring 1998; Equation 2 of Karastergiou \& Johnston 
2004)
\begin{equation}
\langle z\rangle = \sigma_n\sqrt{\frac{2}{\pi}}\exp{\left(-\frac{\mu_v^2}{2\sigma_n^2}\right)}
                   + \mu_v{\rm erf}\left(\frac{\mu_v}{\sigma_n\sqrt{2}}\right),
\label{eqn:Vavg}
\end{equation}
where ${\rm erf}(x)$ is the error function. The threshold value of the mean estimator is 
determined from $\langle z\rangle$ when $\mu_v=0$ and is equal to $K_s=\sqrt{2/\pi}$.

The mode estimator for $|V|$ is derived by setting the derivative of $f_{|V|}(z)$ with respect 
to $z$ equal to zero (Wardle \& Kronberg 1974). The mode estimator is the solution to
\begin{equation}
\frac{z}{\mu_v} = \tanh{\left(\frac{z\mu_v}{\sigma_n^2}\right)}.
\end{equation}
The mode of $f_{|V|}(z)$ occurs at $z=0$ until $\mu_v>\sigma_n$. The threshold value of the 
mode estimator is $K_w=0$.

The median estimator for $|V|$ is the median, $y$, of $f_{|V|}$ (Simmons \& Stewart 1985)
and satisfies the relation
\begin{equation}
1 = {\rm erf}\left(\frac{y+\mu_v}{\sigma_n\sqrt{2}}\right) + 
    {\rm erf}\left(\frac{y-\mu_v}{\sigma_n\sqrt{2}}\right).
\label{eqn:Vmed}
\end{equation}
The threshold value for the median estimator is determined from Equation~\ref{eqn:Vmed} 
when $\mu_v=0$. It is equal to $K_m = {\rm erf}^{-1}(1/2)\sqrt{2} = 0.6745$, where 
${\rm erf}^{-1}(x)$ is the inverse error function.

The ML estimator is found by maximizing $f_{|V|}(z)$ with respect to $\mu_v$ (Simmons \&
Stewart 1985) and satisfies the relation
\begin{equation}
\frac{z}{\mu_v} = \coth{\left(\frac{z\mu_v}{\sigma_n^2}\right)}.
\label{eqn:Vml}
\end{equation}
The threshold value of the ML estimator is $K_{ml} = 1$. The threshold values of all four 
estimators are compiled in the second column of Table 1. 

The estimators are compared in the bottom panel of Figure~\ref{fig:Vdist}. All four 
estimators converge at an SNR of about $s=1.6$, because the pdf of $|V|$ approaches the 
Gaussian pdf of $V$ at intermediate to high SNR. The panel shows that measurements of 
$|V|$ are not biased by the instrumental noise at intermediate to high SNR. The asymptotic 
behavior of the estimators suggests they may be approximated by a simple detection 
threshold:
\begin{equation}
|V|_t =
\begin{cases}
|V|_m & \quad \mathrm{if}\ |V|_m > K_a\sigma_n \\
0 & \quad \mathrm{otherwise}.
\end{cases}
\label{eqn:Vthreshold}
\end{equation}
The approximation to the mode estimator (i.e., when $K_a=K_w=0$) is the AS estimator for 
$|V|$ and is shown by the open circles in the bottom panel of the figure. Since it does not 
compensate measurements for instrumental noise, it is also the naive estimator for $|V|$. 

\begin{figure}
\plotone{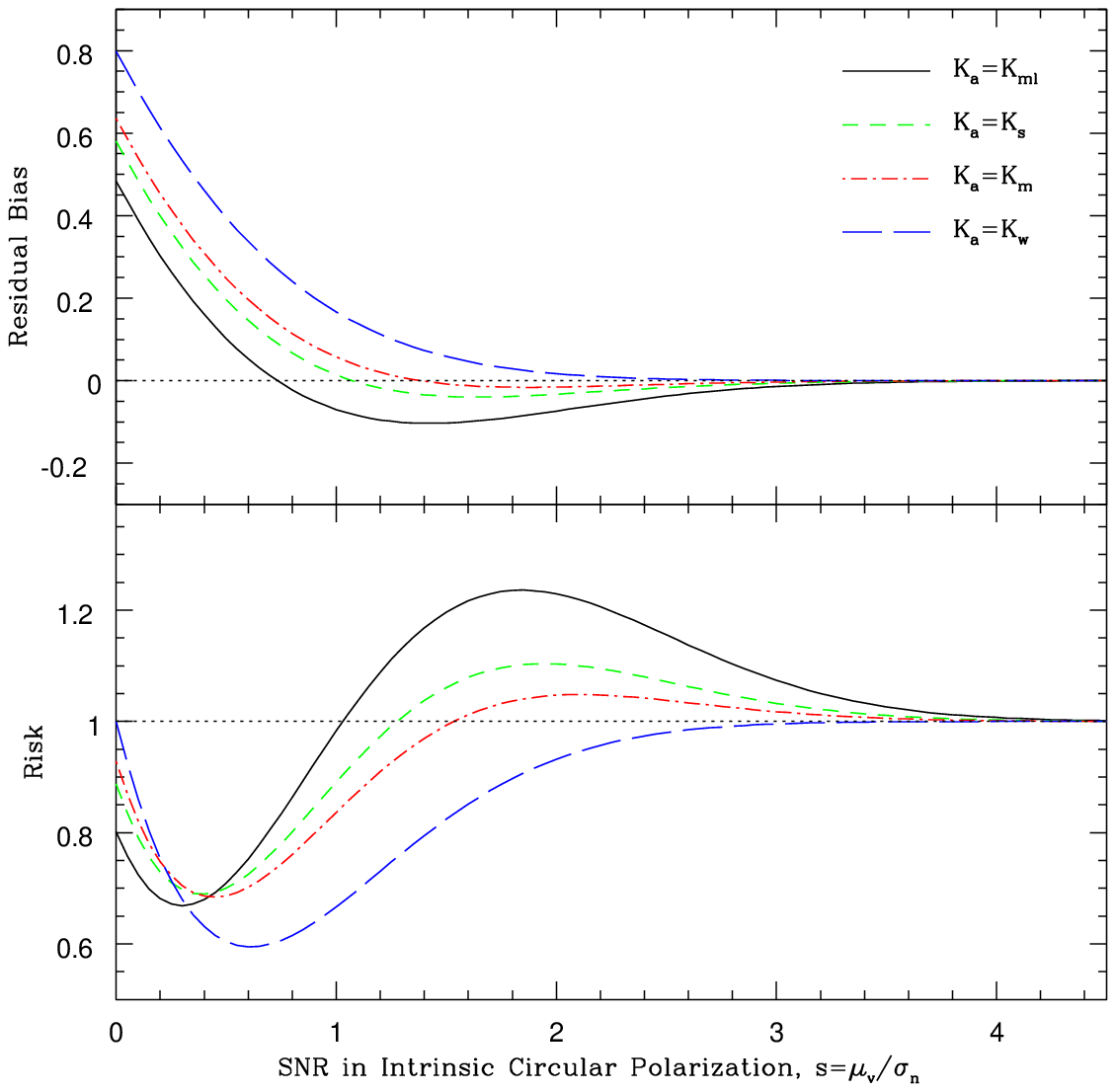}
\caption{Residual bias and risk of the approximations to the circular polarization estimators. 
The top panel compares the residual bias of the approximations calculated from 
Equation~\ref{eqn:Vthreshold} using different threshold values, $K_a$. The bottom panel 
compares the risk of the estimator approximations.}
\label{fig:Vrisk}
\end{figure}

The pdf of $|V|_t$ is comprised of $f_{|V|}(z)$ truncated on its low side at $K_a\sigma_n$ 
and a delta function at $z=0$. The residual bias and risk of the estimator approximations 
can be calculated analytically from the first and second moments of this pdf using their 
respective values of $K_a$. They are compared in Figure~\ref{fig:Vrisk}. The AS estimator 
quickly converges to zero bias but at the expense of a large bias and risk at small SNR. 
The other estimator approximations overshoot zero bias near $s\simeq 1$, but produce a 
lower bias than the AS estimator at $s<1$. Similarly to Simmons \& Stewart's (1985) finding 
for $L$, the approximation to the ML estimator for $|V|$ produces the lowest bias at low 
SNR ($s<0.8$).

\subsection{Total Polarization Estimation}

\begin{figure}
\plotone{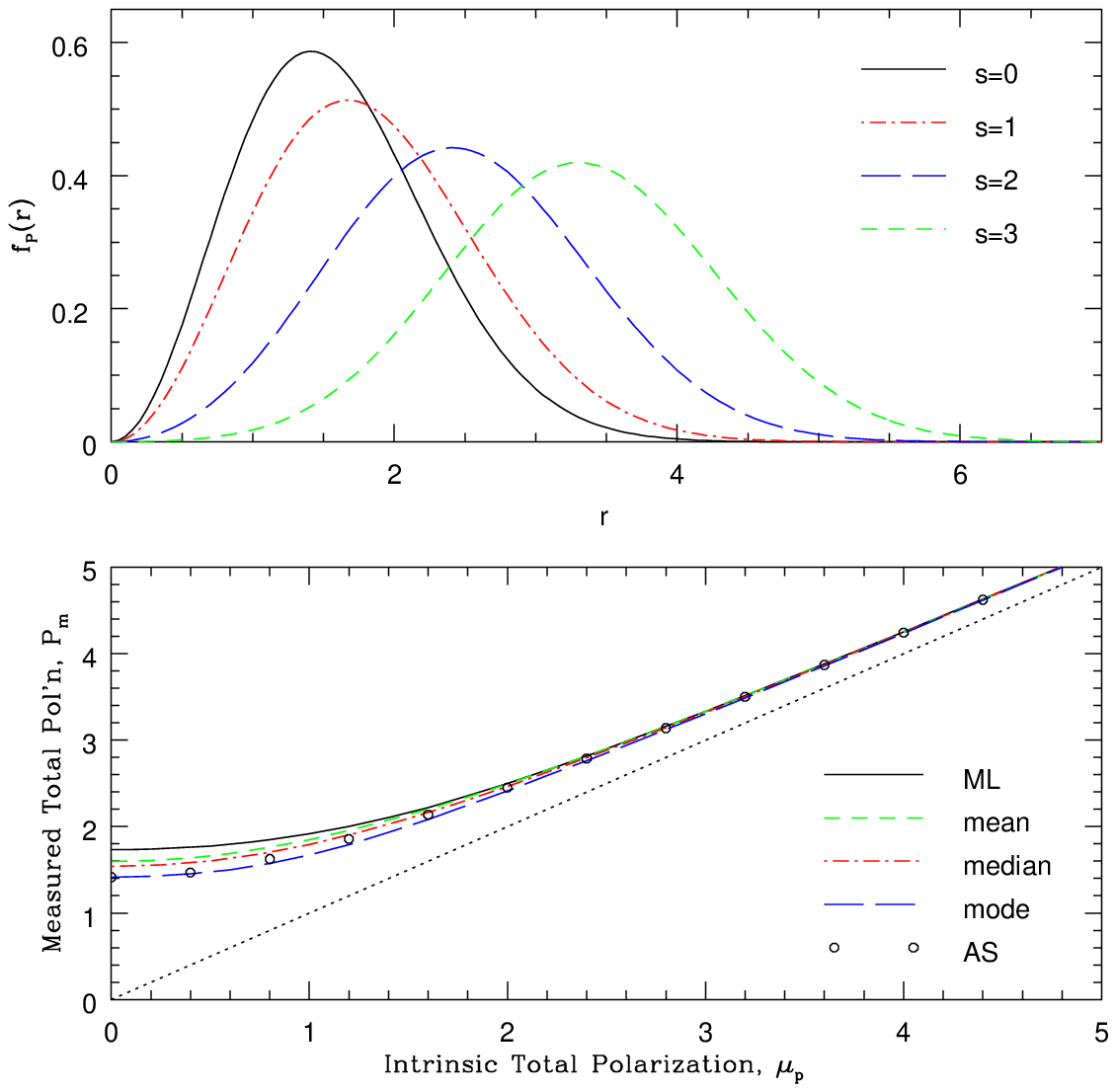}
\caption{Probability density function (pdf) and polarization estimators for total 
polarization when the amplitude of the polarization vector is constant. The top panel 
shows the pdf of $P$ (Equation~\ref{eqn:fixdist}) for different values of SNR in total 
polarization, $s=\mu_p/\sigma_n$. The bottom panel compares the ML, mean, median, mode, 
and AS estimators for $P$. The dotted diagonal line denotes $P_m=\mu_p$.}
\label{fig:Pdist}
\end{figure}

When the amplitude of the total polarization vector is constant at $\mu_p$, the pdf
of the measured vector amplitude is (e.g., Equation 11 of McKinnon 2003; Equation 24 
of Quinn 2014)
\begin{equation}
f_P(r) = \frac{r}{\mu_p\sigma_n}\sqrt{\frac{2}{\pi}}
         \exp{\left[-\frac{(r^2+\mu_p^2)}{2\sigma_n^2}\right]}
         \sinh{\left(\frac{r\mu_p}{\sigma_n^2}\right)}.
\label{eqn:fixdist}
\end{equation}
Examples of $f_P(r)$ are shown in the top panel of Figure~\ref{fig:Pdist} for different 
values of SNR in total polarization, $s=\mu_p/\sigma_n$. The pdf is Maxwell-Boltzmann 
when $s=0$, and is Gaussian-like when $s\gg 1$. Quinn (2014) derived the mode and median 
estimators for $P$, and McKinnon (2003) calculated its mean estimator. All four estimators 
and their threshold values are compiled in Appendix~\ref{sec:Pest}. The threshold values 
are also listed in the last column of Table 1. The estimators are shown in the bottom 
panel of Figure~\ref{fig:Pdist}. The measured polarization, $P_m$, predicted by all four 
estimators converges to $P_m = \sqrt{\mu_p^2 + 2\sigma_n^2}$ at an SNR of about $s=2$ 
(Quinn 2014). The asymptotic behavior of $P_m$ is due to $f_P(r)$ evolving to a Gaussian 
with a mean of $\sqrt{\mu_p^2 + 2\sigma_n^2}$ and a standard deviation of $\sigma_n$. 
The AS estimator for $P$ is then given by Equation~\ref{eqn:approx} with $L$ replaced 
by $P$ and with $K_a=\sqrt{2}$. It is shown by the open circles in the bottom panel of 
Figure~\ref{fig:Pdist}.

\begin{deluxetable}{cccc}
\tablenum{1}
\tablecaption{Estimator Threshold Values, $K_a$}
\tablehead{\colhead{Polarization} & \colhead{Circular, $|V|$} & \colhead{Linear, $L$} &
           \colhead{Total, $P$}} 
\startdata
Mode, $K_w$ & 0 & 1 & $\sqrt{2}$ \\ 
Median, $K_m$ & ${\rm erf}^{-1}(1/2)\sqrt{2}$ & $\sqrt{2\ln(2)}$ & 1.5382 \\
Mean, $K_s$ & $\sqrt{2/\pi}$ & $\sqrt{\pi/2}$ & $\sqrt{8/\pi}$ \\
ML, $K_{ml}$ & 1 & $\sqrt{2}$ & $\sqrt{3}$ \\
\hline
Dimensions, $N$ & 1 & 2 & 3 \\
\enddata
\end{deluxetable}

Trends in the asymptotic behavior and threshold values of the estimators are evident in
their progression from $|V|$ to $L$ to $P$. For each polarization, all estimators converge 
to the same result at high SNR. The threshold values for the mode, median, mean, and ML 
estimators of $|V|$, $L$, and $P$ are compared in Table 1. The threshold values for $L$ 
were derived by Simmons \& Stewart (1985). The table entries are arranged such that the 
threshold values increase down and toward the right of the table. For a specific 
polarization, the mode threshold values are the smallest entries in the table, and the ML 
threshold values are the largest. For each estimator, the threshold for $|V|$ is the 
smallest, and the $P$ threshold is the largest. The last row of the table lists the number 
of dimensions, $N$, from the Poincar\'e sphere used in the derivation of the polarization 
amplitude pdfs. The threshold values of the mode estimator scale as $\sqrt{(N-1)}$ (Quinn 
2014), and the threshold values for the ML estimator scale as $\sqrt{N}$. These trends 
suggest that the basic form of the EW estimator can be generalized to a hybrid estimator 
for $X = |V|$, $L$, or $P$:
\begin{equation}
X_t =
\begin{cases}
  \left(X_m^2 - K_w^2\sigma_n^2\right)^{1/2} & \quad \mathrm{if}\ X_m > K_c\sigma_n \\
  0 & \quad \mathrm{otherwise}.
\end{cases}
\label{eqn:hybrid}
\end{equation}
The constant $K_w$ in Equation~\ref{eqn:hybrid} is the relevant mode threshold from Table 1. 
It accounts for the noise contribution to the measured polarization at high SNR. The cutoff 
value, $K_c$, compensates the measured polarization for instrumental noise at low SNR and 
is constrained by $K_c\ge K_w$ to ensure $X_t$ is real. When $X=|V|$, the hybrid estimator 
becomes the estimator for $|V|$ given by Equation~\ref{eqn:Vthreshold}, because $K_w=0$. 
The hybrid estimator is the AS estimator when $K_c=K_w$ and is the EW estimator when $X=L$ 
and $K_c=1.57$. 

\begin{figure}
\plotone{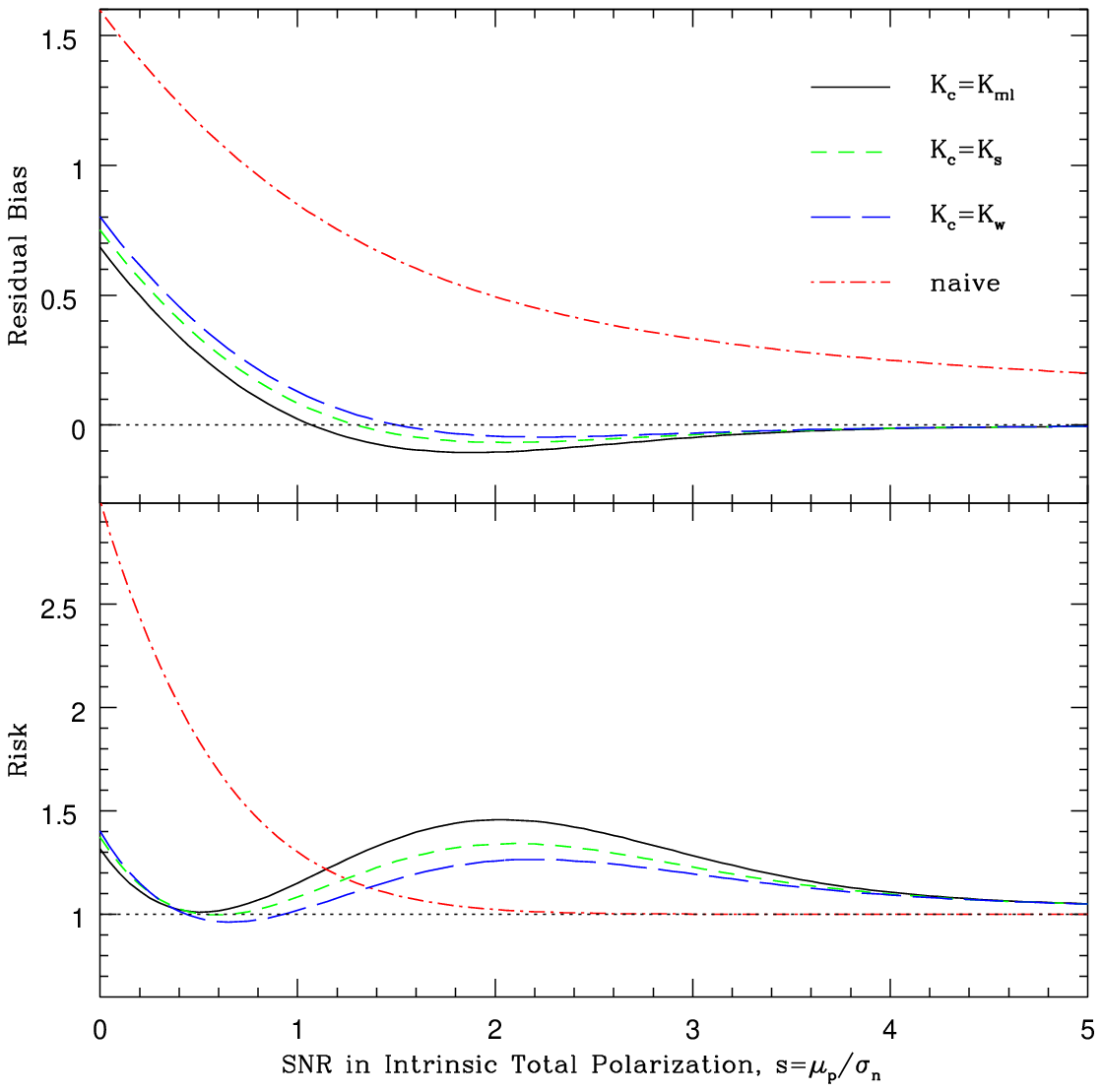}
\caption{Residual bias and risk of the naive and hybrid estimators for total polarization. 
The top panel compares the residual bias of the hybrid estimators calculated from 
Equation~\ref{eqn:hybrid} using $K_w=\sqrt{2}$ and cutoff values of $K_c=K_{ml}$, $K_s$, and 
$K_w$ for $P$ from Table 1.  The bottom panel compares the risk of the estimators.}
\label{fig:Prisk}
\end{figure}

Figure~\ref{fig:Prisk} compares the residual bias and risk of the total polarization estimators 
calculated from Equation~\ref{eqn:hybrid} using $K_w=\sqrt{2}$ and different cutoff values, 
$K_c$, as listed in the figure. The bias and risk of the naive estimator for $P$ were calculated 
analytically from the first and second moments of the pdf given by Equation~\ref{eqn:fixdist}. 
The bias and risk of the hybrid estimators were calculated numerically with Monte Carlo 
simulations. A simulation generated a half million ($2^{19}$) samples each of the Stokes 
parameters $Q$, $U$, and $V$ at intrinsic polarization values ranging from $\mu_p=0$ to $\mu_p=5$ 
at intervals of $\Delta\mu_p=0.05$. The hybrid estimator (Equation~\ref{eqn:hybrid}) was then 
applied to the individual values of $P_m$ calculated from sets of $Q$, $U$, and $V$. The residual 
bias and risk of the resulting values of $P_t$ were calculated from the first and second moments 
of $P_t$. This procedure was repeated 64 times at each $\mu_p$ interval to produce the mean bias 
and risk shown in the figure. The figure shows the significant improvements that can be made in 
bias removal with the hybrid estimator over the naive estimator. The bias of the naive estimator 
is approximately $\sigma_n^2/\mu_p$ at high SNR ($s>3$). In contrast, the bias of the hybrid 
estimators converges to zero at $s\simeq 4$. The bias and risk of the naive estimator at low 
SNR ($s<1$) are much larger than their counterparts determined with the hybrid estimator. Of 
the hybrid estimators, the ML estimator ($K_c=K_{ml}$) produces the lowest bias and risk at 
low SNR, but at the expense of slightly higher bias and risk in comparison to the other 
estimators at intermediate SNR ($1<s<3$). The median hybrid estimator ($K_c=K_m$) is not shown 
in the figure. It produces a bias and risk that are intermediate to those produced by the mode 
and mean hybrid estimators. 


\section{Fluctuations in Polarized Intensity}
\label{sec:random}

The procedure for deriving polarization estimators when the amplitude of the polarization
vector fluctuates is generally identical to that for a vector with constant amplitude 
(Section~\ref{sec:constant}). In the following analysis, the pdfs for $|V|$, $L$, and $P$ 
are derived assuming the fluctuations are along the polarization vector. The term 
\lq\lq fluctuations" refers to those occuring from pulse to pulse at a particular pulse 
longitude, as opposed to variations in the average polarization across a pulse profile. 

Cairns et al. (2001, 2004) showed that the intensity fluctuations at specific pulse 
longitudes of PSRs B0833-45, B0950+08, and B1641-45 generally follow a log-normal pdf, 
and McKinnon (2004) showed the fluctuations near the pulse center of PSR B2020+28
follow a Gaussian pdf. Deriving analytical expressions for the polarizations' observed
pdfs and their estimators from a pulsar-intrinsic, log-normal pdf is not straightforward 
(e.g., Karastergiou \& Johnson 2004). However, analytical expressions for the pdfs and
their estimators can be derived when the fluctuations are assumed to be Gaussian or 
exponential RVs. Three scenarios are investigated. The first scenario assumes the 
pulse-to-pulse fluctuations are Gaussian with a mean amplitude that is greater than 
the amplitude fluctuations. The second scenario also assumes the fluctuations are 
Gaussian, but with a mean amplitude that is equal to zero, in which case the polarized 
signal consists of fluctuations, only. This can occur in pulsars through the incoherent 
addition of OPMs having the same mean intensity (McKinnon 2006). The third scenario 
assumes the intrinsic polarization fluctuations are exponential. Given the larger 
modulation index of exponential intensity fluctuations, which is consistent with 
the median value of $\beta$ indicated by Burke-Spoloar et al. (2012), this scenario 
may be more representative of what is observed in pulsar radio emission than Gaussian 
intensity fluctuations.

\subsection{Gaussian Fluctuations with Nonzero Mean}
\label{sec:norm}

\begin{figure}
\plotone{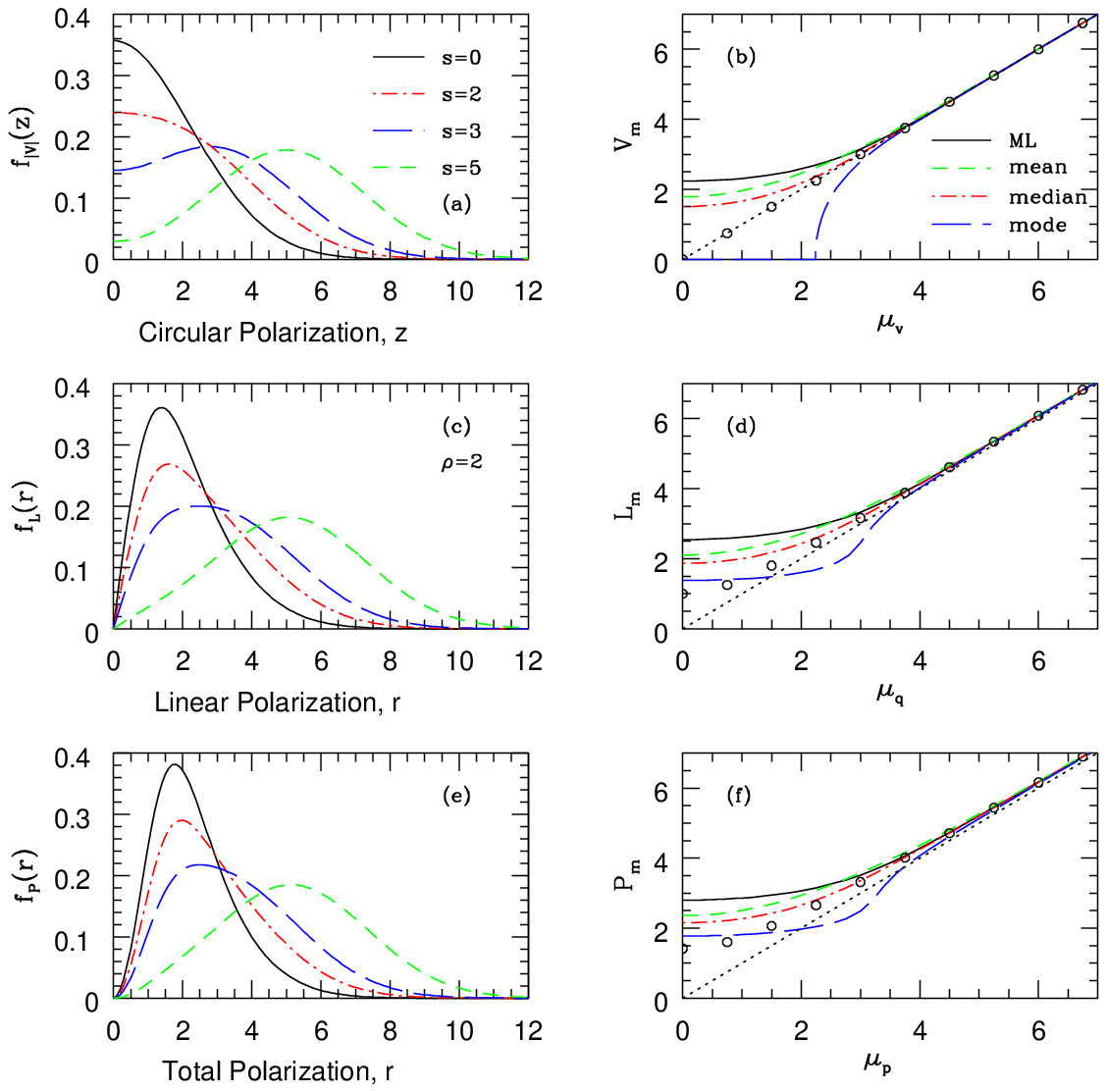}
\caption{Example pdfs and polarization estimators when the polarization fluctuations 
are Gaussian with a nonzero mean. Panels (a), (c), and (e) on the left side of the 
figure show pdfs for $|V|$, $L$, and $P$, respectively, for different values of SNR 
in intrinsic polarization, $s=\mu/\sigma_n$. Panels (b), (d), and (f) on the right 
side of the figure show the ML, mean, median, mode, and AS estimators for $|V|$,
$L$, and $P$, respectively. The open circles denote the AS estimator for each 
polarization. The ratio of the intrinsic polarization fluctuations to the instrumental 
noise in each panel is $\rho=2$.}
\label{fig:normdist}
\end{figure}

When the fluctuations in the Stokes parameter $V$ are Gaussian with a mean of $\mu_v$ and 
a standard deviation of $\sigma_v=\rho_v\sigma_n$, the pdf for $|V|$ is given by 
Equation~\ref{eqn:fv} with $\sigma_n$ replaced by $\sigma_n\sqrt{1+\rho_v^2}$. Similarly, 
the polarization estimators determined from the pdf are the solutions to 
Equations~\ref{eqn:Vavg} -~\ref{eqn:Vml} with $\sigma_n$ replaced by 
$\sigma_n\sqrt{1+\rho_v^2}$. Example pdfs of $|V|$ for different values of SNR in the 
mean polarization, $s=\mu_v/\sigma_n$, and a fixed value of $\rho_v=2$ are shown in 
Figure~\ref{fig:normdist}(a). The polarization estimators for $|V|$ are shown in 
Figure~\ref{fig:normdist}(b).


Assuming the linear polarization is concentrated in the Stokes parameter $Q$ and the fluctuations
in $Q$ are Gaussian with a mean of $\mu_q$ and a standard deviation of $\sigma_q=\rho_q\sigma_n$,
the joint probability density of the amplitude of the linear polarization vector and its
position angle, $\psi$, is (Equation 33 of McKinnon 2003)
%
%
\begin{equation}
f(r,\psi) = \frac{r}{\sigma_n^2\pi(1+\rho_q^2)^{1/2}}
            \exp{\left[-\frac{\mu_q}{2\sigma_n^2(1+\rho_q^2)}\right]}
            \exp{\left\{-\frac{r^2[1+\rho_q^2\sin^2(2\psi)]}{2\sigma_n^2(1+\rho_q^2)}\right\}}
            \exp{\left[\frac{r\mu_q\cos(2\psi)}{\sigma_n^2(1+\rho_q^2)}\right]}.
\label{eqn:Lg}
\end{equation}
The pdf for $L$ can be found by numerically integrating Equation~\ref{eqn:Lg} over $\psi$. 
The polarization estimators for $L$ can then be determined numerically from the resulting 
pdf. Example pdfs of $L$ for different values of SNR in its mean polarization, 
$s=\mu_q/\sigma_n$, and a fixed value of $\rho_q=2$ are shown in Figure~\ref{fig:normdist}(c). 
The polarization estimators for $L$ are shown in Figure~\ref{fig:normdist}(d).


When the fluctuations in total polarization are Gaussian with a mean of $\mu_p$ and a 
standard deviation of $\sigma_p=\rho_p\sigma_n$, the pdf for $P$ from Equation 5 of 
McKinnon (2006) is
\begin{equation}
f_P(r) = 
   \frac{r}{2\rho_p\sigma_n^2}\exp{\left(-\frac{r^2}{2\sigma_n^2}\right)}
   \exp{\left(-\frac{\mu_p^2}{2\rho_p^2\sigma_n^2}\right)}
   \{{\rm erfi}[h_+(r)] - {\rm erfi}[h_-(r)]\},
\label{eqn:fp}
\end{equation}
where ${\rm erfi}(x)=(-i){\rm erf}(ix)$ is the imaginary error function, $i=\sqrt{-1}$, 
and the functions $h_{\pm}(r)$ are given by 
\begin{equation}
h_{\pm}(r) = \frac{\mu_p\pm r\rho_p^2}{\rho_p\sigma_n\sqrt{2(1+\rho_p^2)}}.
\end{equation}
Example pdfs of $P$ for different values of SNR in its mean polarization, $s=\mu_p/\sigma_n$,
and a fixed value of $\rho_p=2$ are shown in Figure~\ref{fig:normdist}(e). 

The mode estimator for $P$ is the solution to
\begin{equation}
\frac{r\rho_p\sigma_n}{r^2-\sigma_n^2} \sqrt{\frac{2}{\pi(1+\rho_p^2)}}
   = \frac{{\rm erfi}[h_+(r)]-{\rm erfi}[h_-(r)]}
    {\exp{\left[h_+^2(r)\right]} + \exp{\left[h_-^2(r)\right]}},
\end{equation}
and the ML estimator satisfies the relation 
\begin{equation}
\frac{\rho_p\sigma_n}{\mu_p}\sqrt{\frac{2}{\pi(1+\rho_p^2)}}
   = \frac{{\rm erfi}[h_+(r)]-{\rm erfi}[h_-(r)]}
    {\exp{\left[h_+^2(r)\right]} - \exp{\left[h_-^2(r)\right]}}.
\end{equation}
The mean and median estimators for $P$ can be calculated numerically from the pdf
given by Equation~\ref{eqn:fp}. The polarization estimators for $P$ are shown in 
Figure~\ref{fig:normdist}(f).

Figure~\ref{fig:normdist} shows that the pdfs for $|V|$, $L$, and $P$ and their estimators
evolve in similar ways. As the SNR increases, the pdfs become Gaussian and the estimators
approach their respective AS estimators. The estimator panels on the right side of the
figure show the constant polarization, $\mu$, is the primary contributor to the measured 
polarization at high SNR, while the polarization fluctuations, $\sigma$, are the main 
contributor to the measured polarization at low SNR.

Consider a hypothetical radio source with Gaussian fluctuations in its total intensity and
a constant polarization fraction. For the intrinsic total intensity to remain nonnegative,
the mean intensity must exceed the fluctuations by about a factor of 5 ($\mu\ge 5\sigma$ or
$\beta < 0.2$; McKinnon 2004). The same is true of the polarization since it is assummed 
to be a constant fraction of the total intensity. If the fluctuations are also greater
than the instrumental noise ($\rho > 1$), the measured polarization of the source will
reside in the upper-right corner of the estimator panels in Figure~\ref{fig:normdist}. A
conclusion that can be drawn from this scenario is the estimators derived for a source 
with a constant polarization amplitude can be used to estimate the mean polarization of 
a source with relatively small fluctuations in its polarization.


\subsection{Gaussian Fluctuations with Zero Mean}
\label{sec:opm}

\begin{figure}
\plotone{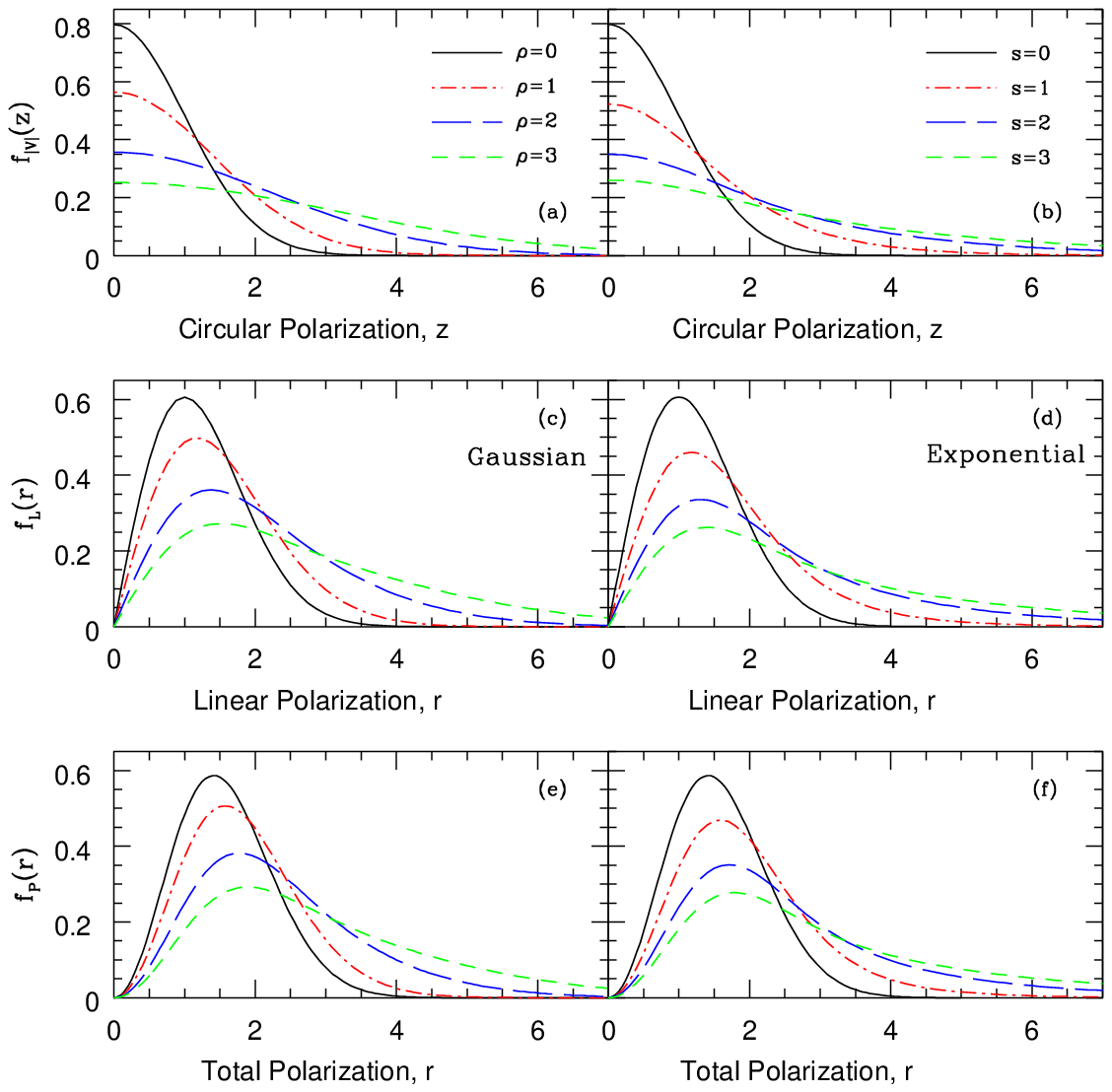}
\caption{Comparison of pdfs for Gaussian and exponential fluctuations in polarization. 
Panels (a), (c), and (e) on the left side of the figure show the pdfs for $|V|$, $L$, 
and $P$, respectively, when the polarization fluctuations are Gaussian. The example
pdfs are shown for different values of the parameter $\rho=\sigma/\sigma_n$ as annotated 
in panel (a). Panels (b), (d), and (f) on the right side of the figure show the pdfs
for $|V|$, $L$, and $P$, respectively, when the fluctuations are exponential. The pdfs
are shown for different values of SNR in intrinsic polarization, $s=\mu/\sigma_n$, as 
annotated in panel (b).}
\label{fig:distcomp}
\end{figure}

The analysis heretofore has focused on an estimator's ability to replicate the constant 
component of the polarization. Now the focus shifts to testing an estimator's ability to 
replicate the polarization's fluctuating component.

\begin{figure}
\plotone{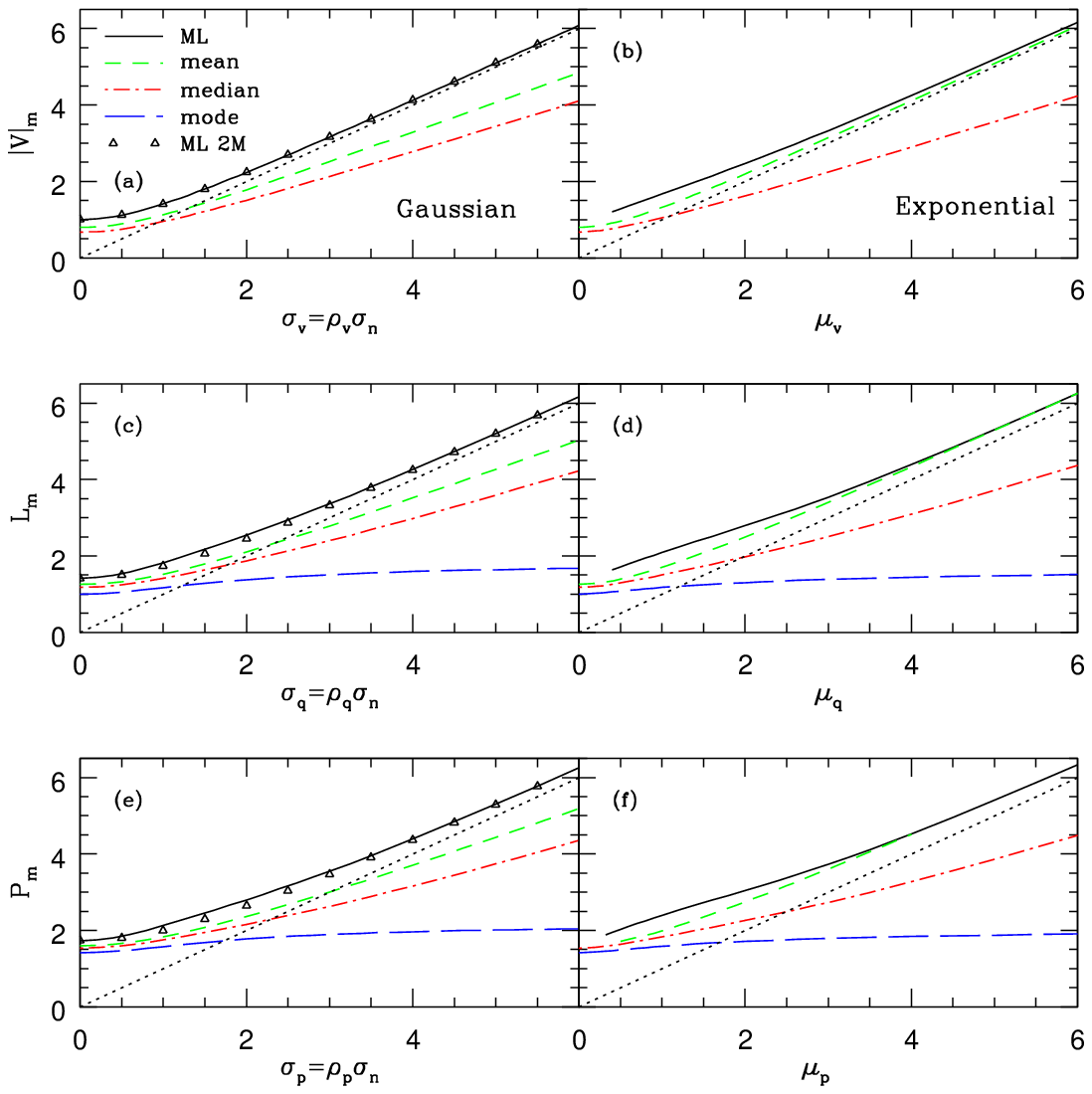}
\caption{Comparison of the polarization estimators for Gaussian and exponential fluctuations 
in polarization. Panels (a), (c), and (e) on the left side of the figure show the ML, mean, 
median, and mode estimators for $|V|$, $L$, and $P$, respectively, when the polarization
fluctuations are Gaussian. The open triangles denote the approximations to the ML estimators 
determined by the polarization's second moment (2M). Panels (b), (d), and (f) on the right 
side of the figure show the estimators for $|V|$, $L$, and $P$, respectively, when the 
fluctuations are exponential. The mode estimators for $|V|$ in panels (a) and (b) are always 
equal to zero.}
\label{fig:estcomp}
\end{figure}

When the fluctuations in the Stokes parameter $V$ are Gaussian with a mean of $\mu_v=0$ 
and a standard deviation of $\sigma_v=\rho_v\sigma_n$, the pdf for $|V|$ is given by 
Equation~\ref{eqn:fv} with $\mu_v$ set to zero and $\sigma_n$ replaced by 
$\sigma_n\sqrt{1+\rho_v^2}$. Example pdfs of $|V|$ for different values of $\rho_v$ are 
shown in Figure~\ref{fig:distcomp}(a). The estimators derived from the pdf are listed in 
Appendix~\ref{sec:OPMest}. The ML estimator is the square root of the second moment of $|V|$, 
and the other estimators are proportional to it. The polarization estimators for $|V|$ are 
shown in Figure~\ref{fig:estcomp}(a).

When the fluctuations in the Stokes parameter $Q$ are Gaussian with a mean of $\mu_q=0$,
the pdf of $L$ is given by Equation 34 of McKinnon (2003):
\begin{equation}
f_L(r) = \frac{r}{\sigma_n^2(1+\rho_q^2)^{1/2}}\exp{\left[-\frac{r^2(2+\rho_q^2)}
         {4\sigma_n^2(1+\rho_q^2)}\right]}
         I_0\left[\frac{r^2\rho_q^2}{4\sigma_n^2(1+\rho_q^2)}\right]
\label{eqn:opmdist}
\end{equation}
The function $I_0(x)$ in the pdf is the modified Bessel function of order zero. Example 
pdfs of $L$ for different values of $\rho_q$ are shown in Figure~\ref{fig:distcomp}(c). The mean, 
mode, and ML estimators for $L$ are listed in Appendix~\ref{sec:OPMest}. The median estimator 
can be calculated numerically from Equation~\ref{eqn:opmdist}. The polarization estimators for 
$L$ are shown in Figure~\ref{fig:estcomp}(c).

The pdf for Gaussian fluctuations only, in total polarization, is given by Equation~\ref{eqn:fp} 
with $\mu_p=0$:
\begin{equation}
f_P(r) = \frac{r}{\rho_p\sigma_n^2}\exp{\left(-\frac{r^2}{2\sigma_n^2}\right)}
              {\rm erfi}\left[\frac{r\rho_p}{\sigma_n[2(1+\rho_p^2)]^{1/2}}\right]
\label{eqn:muzero}
\end{equation}
Example pdfs of $P$ for different values of $\rho_p$ are shown in Figure~\ref{fig:distcomp}(e). 
The mean, mode, and ML estimators for $P$ are listed in Appendix~\ref{sec:OPMest}. The median 
estimator can be calculated numerically from Equation~\ref{eqn:muzero}. The polarization 
estimators for $P$ are shown in Figure~\ref{fig:estcomp}(e).

The pdfs for $|V|$, $L$, and $P$ on the left side of Figure~\ref{fig:distcomp} are generally
similar to one another and evolve in similar ways. The same is true of their respective estimators, 
as shown on the left side of Figure~\ref{fig:estcomp}. The pdfs are generally skewed half-Gaussians 
or asymmetric Gaussians with tails that extend as the fluctuations increase. Unlike the estimators 
for constant polarization that converge at high SNR, the estimators for fluctuating polarization 
diverge as the fluctuations increase. The divergence of the estimators is caused by the increased 
skewness of the pdfs. The quandary posed by the divergence of the estimators is which estimator, 
if any, can be used as a reliable indicator of the intrinsic polarization fluctuations. The mode 
estimator is ineffective at predicting the polarization fluctuations, because it is essentially 
constant for all values of $\rho$. Owing to its large slope, the ML estimator is more sensitive
to polarization fluctuations than the other estimators. As documented in Appendix~\ref{sec:OPMest} 
and as shown by the open triangles in panels (a), (c), and (e) of Figure~\ref{fig:estcomp}, the 
ML estimator for $|V|$, $L$, and $P$ can be approximated by the square root of their second 
moments. The same cannot be said for the ML estimators derived for a polarization vector with 
constant amplitude (see Section~\ref{sec:constant}). Practical estimators for polarization 
fluctuations are determined empirically from Equation~\ref{eqn:hybrid} in Section~\ref{sec:exp}.


\subsection{Exponential Fluctuations }
\label{sec:exp}

As mentioned in Section~\ref{sec:intro} and the introduction to Section~\ref{sec:random}, 
the intrinsic fluctuations in pulsar radio emission may be better represented by an 
exponential RV than a Gaussian RV. Additionally, a complete assessment of a Gaussian RV's 
contribution to the measured polarization requires the independent determination of the 
two free parameters, $\mu$ and $\sigma$, that characterize its pdf. Only a single parameter 
must be determined in the case of an exponential RV, because the mean and standard deviation 
of its pdf are equal. The polarization pdfs and estimators arising from exponential 
fluctuations in the amplitude of a polarization vector are derived in the analysis that 
follows.


When the intrinsic fluctuations in circular polarization are exponential with a mean of 
$\mu_v$, the pdf of the observed $V$ is the convolution of a truncated exponential with 
a zero mean Gaussian that represents the instrumental noise. The resulting pdf is given 
by Equation 4 of McKinnon (2014):
\begin{equation}
f_{\rm V}(z) = \frac{1}{2\mu_v}\exp{\left(\frac{\sigma_n^2}{2\mu_v^2}\right)}
                   \exp{\left(-\frac{z}{\mu_v}\right)}
           \left[{1+{\rm erf}\left(\frac{z-\sigma_n^2/\mu_v}{\sigma_n\sqrt{2}}\right)}\right]
\label{eqn:Vdist}
\end{equation}
The pdf of $|V|$ derived from Equation~\ref{eqn:Vdist} is
\begin{equation}
f_{\rm |V|}(z) = \frac{1}{2\mu_v}\exp{\left(\frac{\sigma_n^2}{2\mu_v^2}\right)}
       \Biggl\{\exp{\left(-\frac{z}{\mu_v}\right)}
       \left[{1+{\rm erf}\left(\frac{z-\sigma_n^2/\mu_v}{\sigma_n\sqrt{2}}\right)}\right]
       + \exp{\left(\frac{z}{\mu_v}\right)}
       {\rm erfc}\left(\frac{z+\sigma_n^2/\mu_v}{\sigma_n\sqrt{2}}\right)\Biggr\}, 
\label{eqn:VXdist}
\end{equation}
%
where ${\rm erfc}(z)$ is the complementary error function. Example pdfs of $|V|$ for different 
values of $s=\mu_v/\sigma_n$ are shown in Figure~\ref{fig:distcomp}(b). The figure shows the 
mode of $f_{|V|}$ always occurs at $z=0$, regardless of the value of $\mu_v$. Therefore, the 
mode estimator for $|V|$ is zero. The mean of $f_{|V|}(z)$ is
\begin{equation}
\langle z\rangle = \sigma_n\sqrt{\frac{2}{\pi}} + \mu_v\exp{\left(\frac{\sigma_n^2}{2\mu_v^2}\right)}
                     {\rm erfc}\left(\frac{\sigma_n}{\mu_v\sqrt{2}}\right).
\label{eqn:Vexpavg}
\end{equation}
The first term in Equation~\ref{eqn:Vexpavg} implies that $\langle z\rangle$ includes a constant
contribution from the instrumental noise. However, the second term in the equation offsets
this contribution as the SNR increases, such that the mean approaches
$\langle z\rangle\simeq\mu_v +\sigma_n^2/2\mu_v$ when $\mu_v/\sigma_n\gg 1$.
 
The median, $y$, of $f_{|V|}(z)$ is the solution to
\begin{equation}
1 = 2{\rm erf}\left(\frac{y}{\sigma_n\sqrt{2}}\right)-\exp{\left(\frac{\sigma_n^2}{2\mu_v^2}\right)}
    \Biggl\{\exp{\left(-\frac{y}{\mu_v}\right)}
    \left[{1+{\rm erf}\left(\frac{y-\sigma_n^2/\mu_v}{\sigma_n\sqrt{2}}\right)}\right]
  - \exp{\left(\frac{y}{\mu_v}\right)}
    {\rm erfc}\left(\frac{y+\sigma_n^2/\mu_v}{\sigma_n\sqrt{2}}\right)\Biggr\}.
\end{equation}
%
%
The ML estimator for $|V|$ can be calculated numerically from Equation~\ref{eqn:VXdist}. The 
polarization estimators for $|V|$ are shown in Figure~\ref{fig:estcomp}(b).


Assuming the linear polarization is concentrated in the Stokes parameter $Q$ and the fluctuations
in $Q$ are exponential with a mean of $\mu_q$, the joint probability density of the amplitude of 
the linear polarization vector and its position angle is 
%
%
\begin{equation}
f(r,\psi)=\frac{r}{\mu_q}\frac{1}{\sigma_n\sqrt{2\pi}}\exp{\left(-\frac{r^2}{2\sigma_n^2}\right)}
          \exp{\left\{\frac{[r\cos(2\psi)-\sigma_n^2/\mu_q]^2}{2\sigma_n^2}\right\}}
\Biggl\{1+{\rm erf}\left[\frac{r\cos(2\psi)-\sigma_n^2/\mu_q}{\sigma_n\sqrt{2}}\right]\Biggr\}.
\label{eqn:Lexp}
\end{equation}
The pdf for $L$ can be found by numerically integrating Equation~\ref{eqn:Lexp} over $\psi$. The 
polarization estimators for $L$ can then be determined numerically from the resulting pdf. Example 
pdfs of $L$ for different values of $s=\mu_q/\sigma_n$ are shown in Figure~\ref{fig:distcomp}(d). 
The polarization estimators for $L$ are shown in Figure~\ref{fig:estcomp}(d).


When the intrinsic fluctuations in total polarization are exponential with a mean of $\mu_p$, the 
pdf for $P$ is
\begin{equation}
f_P(r) = \frac{r}{\mu_p\sigma_n\sqrt{2}}\exp{\left(-\frac{r^2}{2\sigma_n^2}\right)}
         \int_{-w_+(r)}^{w_-(r)}\exp(x^2)[1 + {\rm erf}(x)]dx,
\label{eqn:fpx}
\end{equation}
where the functions $w_\pm(r)$ are given by
\begin{equation}
w_\pm(r) = \frac{r\pm\sigma_n^2/\mu_p}{\sigma_n\sqrt{2}}.
\end{equation}
The estimators for $P$ can be calculated numerically from the pdf given by Equation~\ref{eqn:fpx}. 
Example pdfs of $P$ for different values of $s=\mu_p/\sigma_n$ are shown in 
Figure~\ref{fig:distcomp}(f). The polarization estimators for $P$ are shown in 
Figure~\ref{fig:estcomp}(f).

Figure~\ref{fig:distcomp} shows that the pdfs derived for exponential fluctuations in a 
specific polarization ($|V|$, $L$, or $P$) are very similar to their counterparts derived 
for Gaussian fluctuations. The parameter $\rho$ in the figure is a measure of the Gaussian 
fluctuations in polarization, while the parameter $s$ is a measure of the exponential 
fluctuations. When the parameters are equal ($\rho=s$), the respective pdfs are almost 
identical. The tails of the exponential pdfs are slightly more extended. 
Figure~\ref{fig:estcomp} shows that the estimators for Gaussian and exponential 
fluctuations in a specific polarization are also similar. The estimators for both 
generally diverge as the magnitude of the fluctuations increases, although the mean and 
ML estimators for exponential fluctuations approach one another at large values of $\mu$. 
For a specific polarization, the ML, median, and mode estimators produced by exponential  
fluctuations behave similarly and are roughly equal to their counterparts produced by 
Gaussian fluctuations. The similarities between specific estimators suggest they are 
somewhat insensitive to the statistical character of the fluctuations and that the same 
estimator can be used in both cases.

The analysis in Section~\ref{sec:norm} indicates that estimators for a constant vector 
amplitude can replicate the constant component of Gaussian amplitude fluctuations 
provided the constant component is larger than the fluctuating component. The analyses 
of Gaussian and exponential fluctutations in vector amplitude presented in 
Sections~\ref{sec:opm} and~\ref{sec:exp} suggest that estimators for a specific 
polarization may be insensitive to the statistical character of the fluctuations, 
although the best estimator for pure fluctuations in amplitude has yet to be determined. 
Acknowledging these traits, and with the goal of developing estimators for general use,
the cutoff value, $K_c$, of the hybrid estimator proposed in Equation~\ref{eqn:hybrid} 
was empirically optimized to minimize the residual bias produced by exponential 
fluctuations in vector amplitude. Potential candidates for $K_c$ can be any one of the 
threshold values, $K_a$, listed in Table 1. However, as Everett \& Weisberg (2001) 
have indicated, the optimum value of $K_c$ is not restricted to values of $K_a$. The 
optimization trials began with $K_c=K_s$ and $K_c=K_{ml}$, because the residual bias of 
$|V|$ and $P$ shown in Figures~\ref{fig:Vrisk} and~\ref{fig:Prisk} demonstrates they are 
the most effective at removing instrumental bias at low SNR. Henceforth, the estimator 
with the optimum value of $K_c$ is designated as the general purpose (GP) estimator.

\begin{figure}
\plotone{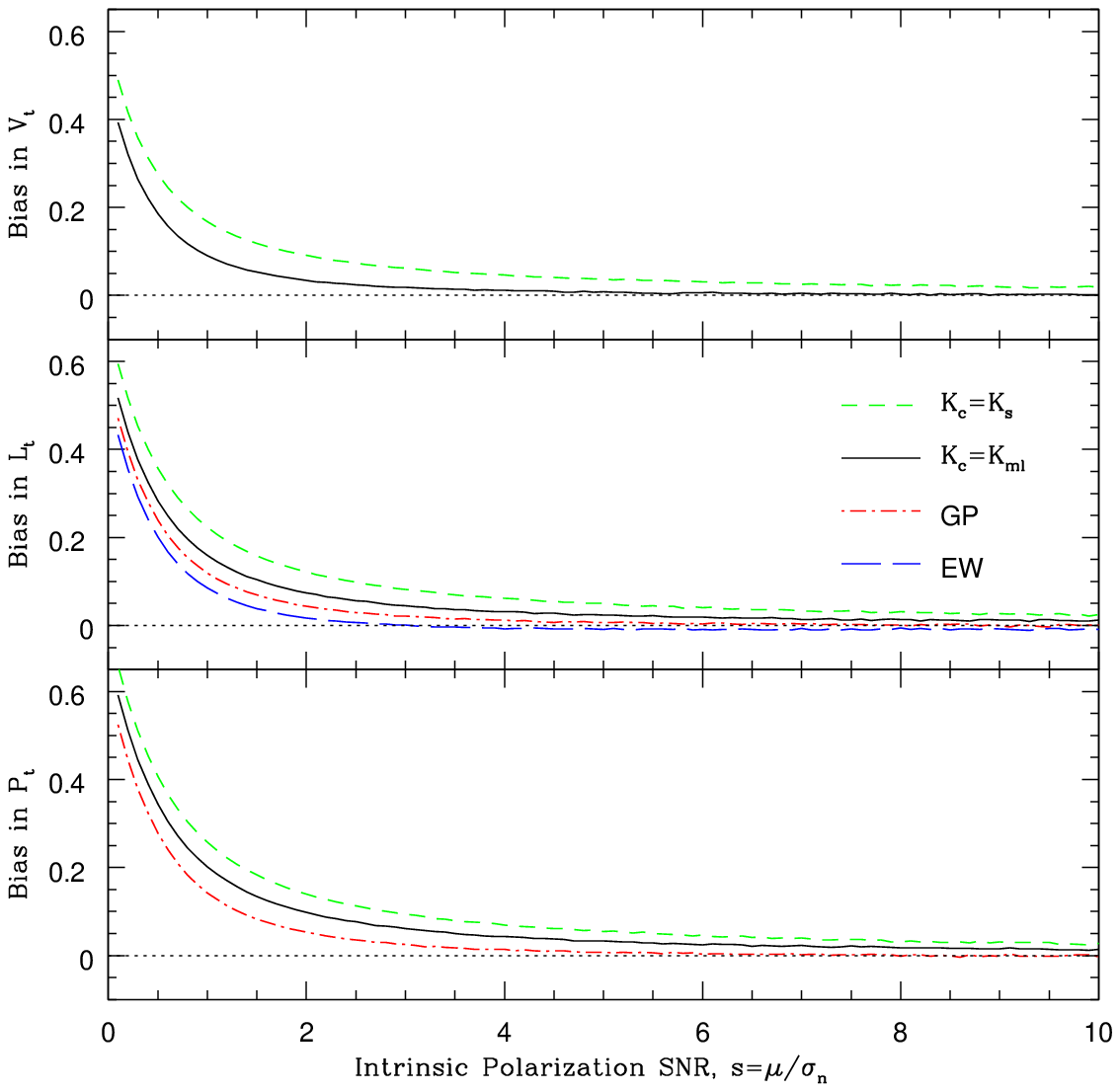}
\caption{Residual bias of the hybrid estimator given by Equation~\ref{eqn:hybrid} for 
$|V|$ (top panel), $L$ (middle panel), and $P$ (bottom panel) when the fluctuations 
in polarization amplitude are exponential. The bias was calculated for different 
cutoff values, $K_c$, and is shown as a function of SNR in the relevant intrinsic 
polarization. The color code annotated in the middle panel of the figure applies to 
all three panels. The parameters $K_w$ and $K_c$ for the GP estimator are listed in 
Table 2. The GP estimator for $|V|$ uses $K_c=K_{ml}$.} 
\label{fig:Xhybrid}
\end{figure}

The results of the optimization are shown in Figure~\ref{fig:Xhybrid}. The top panel of
the figure shows the residual bias in $|V|$ for $K_c=K_s$ and $K_c=K_{ml}$ out to an SNR in 
circular polarization of $s=10$. The estimator is always positively biased for $K_c=K_s$, 
but converges to zero bias at $s\simeq 6$ when $K_c=K_{ml}=1$. Values of $K_c$ larger than 
1 caused the estimator to overshoot zero bias, resulting in a negative bias at high SNR. 
Therefore, the GP estimator for $|V|$ is its hybrid estimator (Equation~\ref{eqn:hybrid})
with $K_c=1$. The middle panel of the figure compares the residual bias produced by 
different values of $K_c$ used with the hybrid estimator for $L$. The values of $K_c$ 
used in the optimization trials are annotated in the panel and include the EW estimator 
with $K_c=1.57$. The panel shows the residual bias of the estimator with $K_c=K_s$ and 
$K_c=K_{ml}$ remains positive through $s=10$. The residual bias of the EW estimator 
becomes slightly negative at $s\simeq 3$. The optimized cutoff value for the GP estimator 
of $L$ was found to be $K_c=1.5$, as shown by the red line in the panel. The bottom panel 
of the figure compares the residual bias produced by different values of $K_c$ used with
the hybrid estimator for $P$. The estimators using $K_c=K_s$ and $K_c=K_{ml}$ are again 
positively biased through $s=10$. The optimized cutoff value for the GP estimator of $P$ 
was found to be $K_c=1.85$. The values of $K_w$ and $K_c$ determined for the GP estimators 
of $|V|$, $L$, and $P$ are summarized in Table 2. In each case, the optimum value of $K_c$ 
is either equal to or slightly greater than $K_{ml}$. Although the parameters for the GP 
estimators were optimized assuming exponential fluctuations in the amplitude of the 
polarization vector, they are consistent with the parameters derived for a vector with 
constant amplitude. The parameters are consistent because the measured polarization 
approaches the intrinsic polarization at high SNR and is dominated by the instrumental 
noise at low SNR in both scenarios. The statistics of the noise are independent of the 
intrinsic polarization fluctuations. Therefore, the GP estimators may be considered for 
general application to pulsar polarization observations.

\begin{deluxetable}{cccc}
\tablenum{2}
\tablecaption{Parameters $K_w$ and $K_c$ for the General Purpose Estimators}
\tablehead{\colhead{Polarization} & \colhead{Circular, $|V|$} & \colhead{Linear, $L$} &
           \colhead{Total, $P$}} 
\startdata
$K_w$ & 0 & 1 & $\sqrt{2}$ \\ 
$K_c$ & 1.0 & 1.5 & 1.85 \\
\enddata
\end{deluxetable}


\section{Comparisons with Other Polarization Estimators}
\label{sec:compare}

The estimator developed by Karastergiou et al. (2003) and Karastergiou \& Johnston (2004)
is often used to compensate measurements of $|V|$ for instrumental noise (e.g., Johnston 
\& Kerr 2018; Serylak et al., 2021; Oswald et al. 2023a; Posselt et al., 2023). The 
estimator is given by (e.g., Equation 6 of Posselt et al. 2023)
\begin{equation}
|V|_t =
\begin{cases}
  |V|_m - \sigma_n\sqrt{2/\pi} & \quad \mathrm{if}\ |V|_m > \sigma_n\sqrt{2/\pi} \\
  0 & \quad \mathrm{otherwise}.
\end{cases}
\label{eqn:KJ}
\end{equation}
The estimator assumes the contribution of the instrumental noise to $|V|_m$ is a constant 
and is essentially independent of the SNR in $V$. However, as shown in 
Figure~\ref{fig:Vdist}, this assumption is not correct because $|V|_m$ and $\mu_v$ 
converge at an SNR of $s\simeq 1.6$. The estimator overcompensates $|V|_m$ for noise such 
that the resulting values of $|V|_t$ are systematically low by $\sigma_n\sqrt{2/\pi}$ for
$s > 1.6$. Values of $|V|_t$ derived from Equation~\ref{eqn:KJ} have also been used to 
calculate the ellipticity angle of a polarization vector, $\chi=0.5\arctan(|V|_t/L_t)$ 
(e.g., Oswald et al. 2023b). Since the estimator produces a value of $|V|_t$ that is 
systematically low, the values of $\chi$ calculated from it are also systematically low. 

Tiburzi et al. (2013) use Equation~\ref{eqn:hybrid} with $X=|V|$, $K_w=\sqrt{2/\pi}$, and 
$K_c=2$ as an estimator for $|V|$. Similarly to Everett \& Weisberg (2001), they do not mention
why they chose their particular value of $K_c$. Their estimator generally overcompensates 
the measurements for instrumental noise and consequently produces estimates of $|V|$ that 
are systematically low by approximately $\sigma_n^2/(\pi\mu_v)$ at high SNR.

\begin{figure}
\plotone{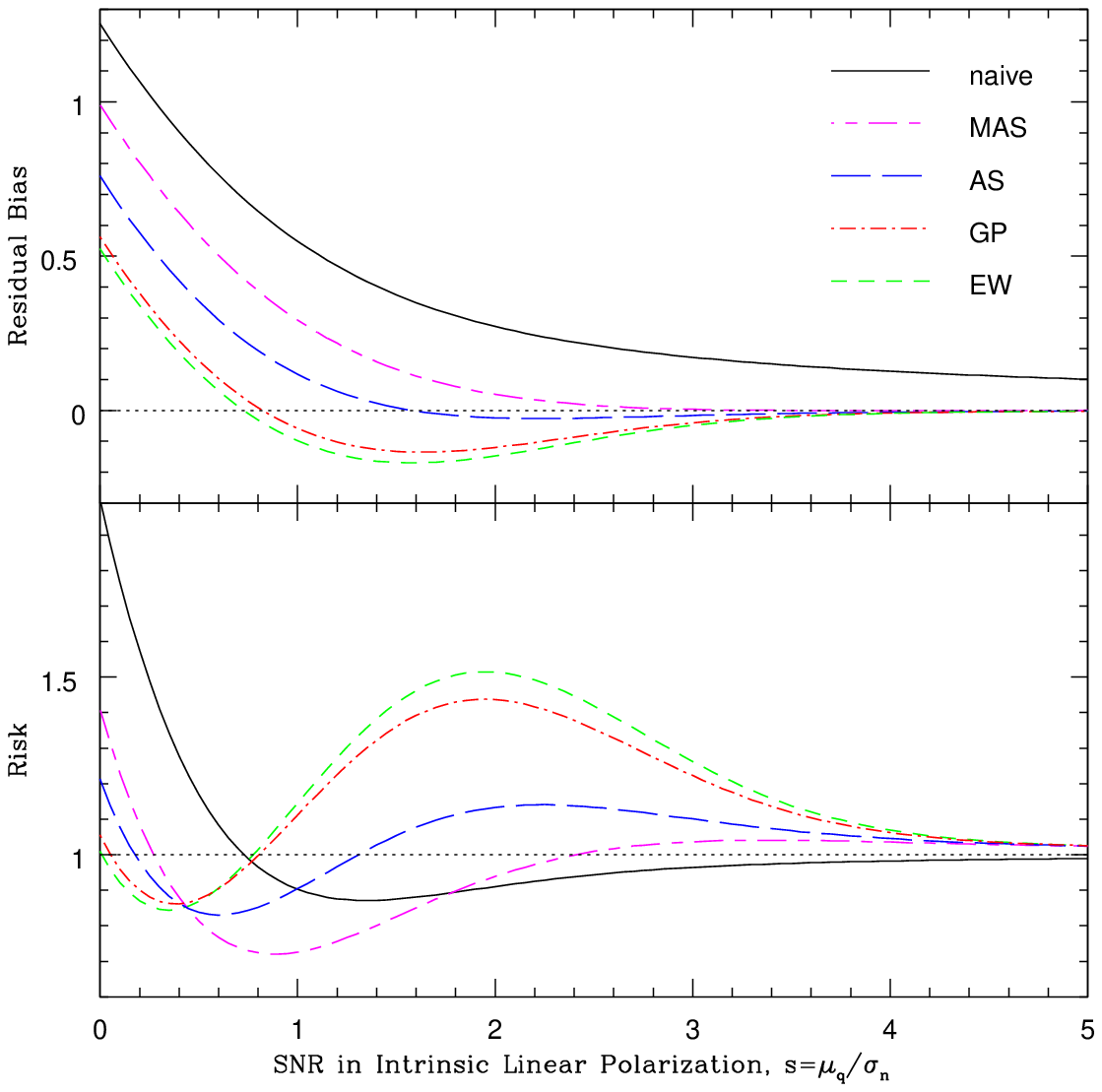}
\caption{Comparison of the residual bias (top panel) and risk (bottom panel) of the naive, 
MAS, AS, GP, and EW estimators when the amplitude of the linear polarization vector is 
constant.}
\label{fig:EWrisk}
\end{figure}

Figure~\ref{fig:EWrisk} compares the residual bias and risk of the naive, MAS, AS, GP, and 
EW estimators for $L$ when the amplitude of the polarization vector is constant. The residual 
bias and risk of the naive estimator shown in the figure were determined analytically from 
the first and second moments of the Rice pdf, and the bias and risk of the other estimators 
were calculated numerically with Monte Carlo simulations. The figure demonstrates the 
improvements that can be made in polarization estimation with the MAS, AS, GP and EW
estimators over no noise compensation (the naive estimator). The MAS, AS, GP, and EW
estimators have little to no residual bias for an SNR greater than $s\simeq 3$, because all 
four estimators have the same asymptotic behavior in the high SNR regime. A comparison 
between the residual bias of the AS, GP, and EW estimators illustrates the effect of their 
different cutoff values. The magnitude of the bias for the GP and EW estimators at low SNR
is smaller than that for the AS estimator, because their larger cutoff values set a larger 
number of data points equal to zero. Therefore, the GP and EW estimators are more effective 
than the AS estimator at removing off-pulse instrumental noise. The trade-off in using the 
GP and EW estimators is their comparatively larger bias and risk at intermediate SNRs. 

The MAS estimator for $L$ is given by Equation 18 of Plaszczynski et al. (2014) and is
replicated in Equation~\ref{eqn:MAS} below. As shown in Figure~\ref{fig:EWrisk}, it 
produces a residual bias that quickly converges to zero, and its risk is generally low 
in comparison to that of other estimators for $s>2$ (see also Figure 3 of Montier 
et al., 2015). However, the compromise made in using the MAS estimator is its large 
residual bias at low SNR where its magnitude is comparable to the instrumental noise. 
The relatively large bias is not unexpected, because the performance of both the AS 
and MAS estimators is optimized for intermediate to high SNR, as their names imply. 
When applied to exponential fluctuations in $L$, the MAS estimator returns a positive 
bias that is in excess of what is shown for the other estimators in the middle panel 
of Figure~\ref{fig:Xhybrid} (e.g., the bias at $s=10$ is approximately 0.07). The MAS 
estimator can also be generalized for application to $X=L$ or $P$ with
\begin{equation}
 X_t = X_m - \frac{K_w^2}{2}\frac{\sigma_n^2}{X_m}
      \left[1-\exp{\left(-\lambda\frac{X_m^2}{\sigma_n^2}\right)}\right].
\label{eqn:MAS}
\end{equation}
The constant $K_w$ in Equation~\ref{eqn:MAS} is the relevant mode threshold from Table 1. 
The constant $\lambda$ is determined empirically and is constrained by $0<\lambda\le 2/K_w^2$ 
to ensure $X_t\ge 0$. Plaszczynski et al. (2014) found that $\lambda=1$ was optimum for 
measurements of $L$. Simulations of the MAS estimator's residual bias indicate that a value 
of $\lambda=3/4$ is suitable for measurements of $P$. The MAS estimator could be applied to 
$|V|$, but it is equivalent to the naive estimator. Consequently, it is not an improvement 
upon the estimators given by Equations~\ref{eqn:Vthreshold} and~\ref{eqn:hybrid}. 

Different estimators have been used to compensate total polarization measurements for 
instrumental noise. For example, Oswald et al. (2023a) and Dial et al. (2025) apply the 
EW estimator for linear polarization to both $L$ and $P$, and Dyks et al. (2021) apply 
Wardle \& Kronberg's (1974) mode estimator for $L$ to both $L$ and $P$. The application 
of estimators for $L$ to measurements of $P$ is not ideal, because the instrumental noise 
contributes more to $P$ than it does to $L$. More specifically, the statistical properties 
of the noise for $L$ and $P$ are different: the noise in $L$ follows a Rayleigh pdf, while 
the noise in $P$ follows a Maxwell-Boltzmann pdf. The EW and mode estimators for $L$ 
undercompensate measurements of $P$ for instrumental noise such that the resulting values 
of $P_t$ are systematically high by approximately $\sigma_n^2/(2\mu_p)$ at high SNR.

Edwards \& Stappers (2004) and MacQuart et al. (2012) note, though do not necessarily recommend,
that a polarization's second moment could be used as a polarization estimator. The attractive
features of the second moment are it captures both the constant and fluctuating components 
of the polarization, its application is independent of the statistical character of the 
polarization fluctuations, and theoretically it can produce a zero residual bias. However, the 
second moment accurately measures the contribution of the instrumental noise only on average 
(MacQuart et al. 2012), and thus may be applicable only when a large number of data samples 
are used in computing the average polarization. The application of the second moment as a
polarization estimator generally overcompensates the measurements for instrumental noise.


\section{SUMMARY}
\label{sec:summary}

Polarization estimators were derived for cases when the amplitude of a polarization vector
is a constant or a RV. The residual bias and risk resulting from the application of 
approximations to the estimators were quantified. A hybrid estimator based on the EW 
estimator for linear polarization was proposed for general application to pulsar polarization 
observations. The parameterization of the hybrid estimators was optimized to minimize their 
residual bias. The optimized hybrid estimators were shown to be more effective at removing 
instrumental noise than their commonly used counterparts.


\acknowledgments{The National Radio Astronomy Observatory and Green Bank Observatory are 
facilities of the U.S. National Science Foundation operated under cooperative agreement 
by Associated Universities, Inc.}


\appendix

\section{Total Polarization Estimators}
\label{sec:Pest}

For a total polarization vector with constant amplitude, the mean estimator for $P$ is given 
by Equation 12 of McKinnon (2003):
\begin{equation}
\langle r\rangle = \sigma_n{\left[\sqrt{\frac{2}{\pi}}\exp{\left(-\frac{s^2}{2}\right)} 
                 + \frac{s^2+1}{s}{\rm erf}\left(\frac{s}{\sqrt{2}}\right)\right]},
\label{eqn:3Damp}
\end{equation}
where $s=\mu_p/\sigma_n$ is the SNR in intrinsic total polarization. The mean approaches 
$\langle r\rangle\simeq\mu_p+\sigma^2/\mu_p$ when $s\gg 1$. The threshold value of the 
mean estimator for $P$ is $K_s = \sqrt{8/\pi}$. 

The median estimator for $P$ is the solution to
%
%
\begin{equation}
1 = {\rm erf}\left(\frac{y+\mu_p}{\sigma_n\sqrt{2}}\right)
  + {\rm erf}\left(\frac{y-\mu_p}{\sigma_n\sqrt{2}}\right)
  - \frac{\sigma_n}{\mu_p}\sqrt{\frac{8}{\pi}}\exp{\left[-\frac{(y^2+\mu_p^2)}{2\sigma_n^2}\right]}
    \sinh{\left(\frac{y\mu_p}{\sigma_n^2}\right)}.
\end{equation}
The threshold value of the median estimator was calculated numerically, and is equal to $K_m=1.5382$. 

From Equation 25 of Quinn (2014), the mode estimator for $P$ is the solution to 
\begin{equation}
\frac{r\mu_p}{r^2-\sigma_n^2} = \tanh{\left(\frac{r\mu_p}{\sigma_n^2}\right)}.
\label{eqn:fixmode}
\end{equation}
The threshold value for the mode estimator is $K_w = \sqrt{2}$.

From Equation 26 of Quinn (2014), the ML estimator for $P$ is the solution to 
\begin{equation}
\frac{r\mu_p}{\mu_p^2+\sigma_n^2} = \tanh{\left(\frac{r\mu_p}{\sigma_n^2}\right)}.
\label{eqn:fixml}
\end{equation}
The threshold value for the ML estimator is $K_w = \sqrt{3}$. 

\section{Estimators for Gaussian Fluctuations Only}
\label{sec:OPMest}

\subsection{Circular Polarization}

When the fluctuations in the Stokes parameter $V$ are Gaussian with a mean of $\mu_v=0$, 
the mode of $f_{|V|}(z)$ always occurs at $z=0$ (see Figure~\ref{fig:distcomp}(a)). 
Therefore, the mode estimator for $|V|$ is zero. The mean estimator for $|V|$ is given 
by Equation~\ref{eqn:Vavg} with $\mu_v$ set to zero and with $\sigma_n$ replaced by 
$\sigma_n\sqrt{1+\rho_v^2}$:
\begin{equation}
\langle z\rangle =\sigma_n\left[2(1+\rho_v^2)/\pi\right]^{1/2}
\end{equation}
Similarly, the median estimator from Equation~\ref{eqn:Vmed} is 
\begin{equation}
y =\sigma_n{\rm erf}^{-1}(1/2)\left[2(1+\rho_v^2)\right]^{1/2}.
\end{equation}
Since $\mu_v=0$, the estimators are tasked with quantifying the fluctuations in polarization 
amplitude. Therefore, the ML estimator is calculated by maximizing the pdf for $|V|$ with 
respect to $\rho_v$. It is given by
\begin{equation}
z_{ml} =\sigma_n(1+\rho_v^2)^{1/2}
\end{equation}
and is equal to the square root of the second moment of $|V|$ (see Figure~\ref{fig:estcomp}(a)).
The mean and median estimators are proportional to the ML estimator.


\subsection{Linear Polarization}

When the fluctuations in the Stokes parameter $Q$ are Gaussian with a mean of $\mu_q=0$, the mean 
estimator for $L$ is the mean of Equation~\ref{eqn:opmdist}:
\begin{equation}
\langle r\rangle = \frac{2\sigma_n(1+\rho_q^2)\sqrt{\pi}}{(2+\rho_q^2)^{3/2}}
                   {}_2F_1[3/4,5/4;1; \rho_q^4/(2+\rho_q^2)^2]
\label{eqn:OPM2D}
\end{equation}
The function ${}_2F_1(\mathrm{a,b;c;} z)$ in Equation~\ref{eqn:OPM2D} is the Gauss hypergeometric 
function with numerical parameters a, b, and c and argument $z$. It is equal to one when $z=0$. 

The mode estimator for $L$ is the solution to
\begin{equation}
\frac{r^2\rho_q^2}{r^2(2+\rho_q^2) - 2\sigma_n^2(1+\rho_q^2)} = 
\frac{I_0[g(r)]}{I_1[g(r)]},
\end{equation}
where the argument of the Bessel functions has been abbreviated by
\begin{equation}
g(r) = \frac{r^2\rho_q^2}{4\sigma_n^2(1+\rho_q^2)}.
\end{equation}
The ML estimator satisfies
\begin{equation}
\frac{r^2}{2\sigma_n^2(1+\rho_q^2)} = 
\frac{I_0[g(r)]}{I_0[g(r)] + I_1[g(r)]}.
\end{equation}
As shown in Figure~\ref{fig:estcomp}(c), the ML estimator can be approximated by the square root 
of the second moment of $L$, $r = \sigma_n(2 + \rho_q^2)^{1/2}$.


\subsection{Total Polarization}

When the fluctuations in total polarization are Gaussian with a mean of $\mu_p=0$, the mean 
estimator for $P$ is the mean of Equation~\ref{eqn:muzero}:
\begin{equation}
\langle r\rangle = \frac{4\sigma_n}{\left[2\pi(1+\rho_p^2)\right]^{1/2}}
                   {}_2F_1[1/2,2;3/2;\rho_p^2/(1+\rho_p^2)]
\label{eqn:OPM3D}
\end{equation}
Alternatively, the mean of $P$ can be written in terms of the inverse hyperbolic tangent,
${\rm arctanh}(x)$:
\begin{equation}
\langle r\rangle = \sigma_n\sqrt{\frac{2}{\pi}}\Biggl\{(1+\rho_p^2)^{1/2}
        + \frac{1}{\rho_p}{\rm arctanh}\left[\frac{\rho_p}{(1+\rho_p^2)^{1/2}}\right]\Biggr\}.
\end{equation}

The mode polarization estimator satisfies the relation
\begin{equation}
\frac{r\rho_p\sigma_n}{r^2-\sigma_n^2}\sqrt{\frac{2}{\pi(1+\rho_p^2)}}
        = {\rm erfi}\left[{\frac{r\rho_p}{\sigma_n\sqrt{2(1+\rho_p^2)}}}\right]
          \exp{\left[-\frac{r^2\rho_p^2}{2\sigma_n^2(1+\rho_p^2)}\right]},
\end{equation}
and the ML estimator is the solution to
\begin{equation}
\frac{r}{\sigma_n}\sqrt{\frac{2}{\pi}}\frac{\rho_p}{(1+\rho_p^2)^{3/2}}
        = {\rm erfi}\left[{\frac{r\rho_p}{\sigma_n\sqrt{2(1+\rho_p^2)}}}\right]
          \exp{\left[-\frac{r^2\rho_p^2}{2\sigma_n^2(1+\rho_p^2)}\right]}.
\end{equation}
As shown in Figure~\ref{fig:estcomp}(e), the ML estimator can be approximated by the square
root of the second moment of $P$, $r = \sigma_n(3 + \rho_p^2)^{1/2}$.


\end{document}